\documentclass[preprint,showpacs,preprintnumbers,amsmath,amssymb]{revtex4}

\usepackage{graphicx}
\usepackage{dcolumn}
\usepackage{bm}

\begin{document}

\title{Effect of density dependent symmetry energy on Elliptical flow.\\}

\author{Suneel Kumar}
\email{suneel.kumar@thapar.edu}
\author{Karan Singh Vinayak}%

\affiliation{%
School of Physics and Materials Science, Thapar University, Patiala-147004, Punjab (India)\\
}%

\date{\today}





\maketitle
\section{Introduction}

Nuclear Physics in general and heavy-ion collisions in particular is of central interest in investigating  the matter under the extreme conditions of temperature and compression.  Among the different observables, collective flow and its various forms enjoys a special status, due to its sensitivity towards model ingredients that define equation of state. The elliptical flow, which is the squeeze out of spectator matter out of the reaction plane,  is more suited to study the collective flow,, as compared to the directed flow, which is constrained only along the reaction plane.\\

 Elliptical flow which is highly sensitive towards symmetry energy, provides us the opportunity to study the pressure which is generated very early in the reaction. The form and strength of the symmetry energy is one of the hot topics these days. The term symmetry energy which is taken as 32 MeV, implies an estimate of the energy cost to convert all the protons to neutrons in nuclear matter at a fixed density of  $0.16 fm^{-3}$. So it is an important goal of the heavy-ion physics to extract information about the symmetry energy of the nuclear matter at densities higher and lower than the  normal nuclear matter density.
The equation below gives us the theoretical conjecture of how the symmetry energy varies against $\rho$\cite{bao}.

\begin{eqnarray}
E(\rho)=E(\rho_o)(\rho/\rho_o)^{\gamma}
\end{eqnarray}

$\gamma$ tells us about the stiffness of the symmetry energy\cite{bao}.
Aim of the present study is to pin down the Elliptical flow for the various forms of the density dependent symmetry energy.\\

\section{ISOSPIN-dependent QUANTUM MOLECULAR DYNAMICS (IQMD) MODEL}

Our calculations are carried out within the framework of Isospin dependent Quantum Molecular Dynamics (IQMD) model which is an extended
form of the QMD model \cite{S.K.R.K.}. The isospin quantum molecular model (IQMD) \cite{S.K.R.K.} is a semi-classical model
which describes the heavy-ion collisions on an event by event basis. For more details, see ref. \cite{S.K.R.K.}.\\
In IQMD model, the centroid of each nucleon propagates under the classical equations of motion \cite{S.K.R.K.}
\begin{equation}
\frac{d\vec{r_i}}{dt}~=~\frac{d\it{H}}{d\vec{p_i}}~~;~~\frac{d\vec{p_i}}{dt}~=~-\frac{d\it{H}}
{d\vec{r_i}}~~\cdot
\end{equation}
The $H$ referring to the Hamiltonian reads as:
\begin{equation}
~H~=~\sum_{i}p_i^2/2m_i~+~V^{ij}_{Yukawa}~~+~V^{ij}_{Coul}~~+~V^{ij}_{skyrme}
+~V^{ij}_{symm}\cdot
\end{equation}\\

\section{Preliminary Results}

For the present analysis simulations are carried out for several  thousand events for the system ${{Au}^{197}_{79}}$ + ${{Au}^{197}_{79}}$  at the incident energy of 100 MeV/nucleon, for the different parametrizations of the density dependence of the symmetry energy. The elliptical flow is defined as the average difference between the square of the x and y components of the transverse momentum. Mathematically, it can be written as\cite{ellip},

\begin{equation}
~v_2~=~\langle~\frac{p_x^2-p_y^2}{p_x^2+p_y^2}~\rangle~\cdot
\end{equation}

\begin{figure}
\includegraphics[scale=0.40]{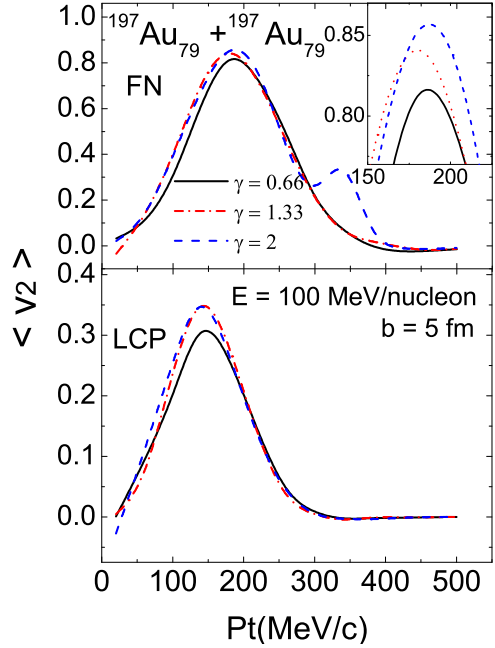}
\caption{\label{fig:1}  Transverse momentum dependence of elliptical flow, summed over the entire rapidity distribution, at b = 5 fm, for different parametrizations of the density dependence of the symmetry energy. The top and bottom panel represents the free nucleons(FN's) and light charged particles(LCP's) respectively .
}
\end{figure}

where $p_{x}~and~p_{y}$ are the x and y components of the momentum. The $p_{x}$ is in the reaction plane and $p_{y}$ is perpendicular to the reaction plane. 
In fig.1, we display the final state elliptical flow for the free particles(upper panel), light charged particles(LCP's)[2~$\leq$~A~$\leq$~4](lower panel) as a function of transverse momentum$(P_{t})$, for the different parametrizations of the density dependence of symmetry energy i.e. $32(\rho/\rho_o)^{\gamma}$, for $\gamma$= 0.66, 1.33 and 2 respectively. This elliptical flow is integrated over the entire rapidity range. Gaussian-type, behaviour is observed for all the forms of the density dependence of the symmetry energy i.e. for the different values of gamma.\\

The effect of symmetry energy is clearly visible in the figure. At the incident energy of 100 MeV/nucleon both the mean field and NN(nucleon-nucleon) collision contributes towards the dynamics of the reaction. The elliptical flow seems to be more sensitive towards the various forms of the density dependence of the symmetry energy for the light charged particles as compared to the free nucleons. The trends observed through our simulations shows the weaker squeeze out flow for the larger values of gamma. The very stiff form of symmetry energy i.e.  $\gamma$ = 2, corresponds to the maximum positive value of elliptical flow. Indeed, in case of LCP, the very stiff symmetry energy i.e. $\gamma$ = 1.33 and 2 does not seems to affect the squeeze out flow on a large scale.



\end{document}